\documentstyle[12pt]{article}
\newcommand{\zero}{\setcounter{equation}{0}}

\begin{document}

\begin{center}

{\Large {\bf Cosmic String Configuration in the Supersymmetric
CSKR Theory}}

\vspace{.80 true cm}

C.N. Ferreira$^a$ \footnote{crisnfer@cbpf.br},
J.A. Helay\"el-Neto $^{a,b}$ \footnote{helayel@gft.ucp.br} and
M.B.D.S.M. Porto$^{a,c}$ \footnote{beatriz@cbpf.br}

\vspace{.50 true cm}

$^a${\it Centro Brasileiro de Pesquisas F\'{\i}sicas (CBPF-CCP), Rua Dr. Xavier Sigaud 150,
Urca 22.290-180, Rio de Janeiro, RJ, Brazil}

\vspace{.25 true cm}

$^b${\it Universidade Cat\'olica de
Petr\'opolis (UCP-GFT), Rua Bar\~ao do Amazonas 124,
25.685-070, Petr\'opolis, RJ, Brazil}

\vspace{.25 true cm}

$^c${\it Departamento de F\'{\i}sica da Universidade Federal Rural
do Rio de Janeiro (UFRRJ), BR465 - KM7, 23.890-000, Serop\'edica, RJ, Brazil}

\end{center}

\vspace{.65 true cm}

\begin{abstract}
We study a cosmic string solution of an  N=1-supersymmetric version of the
Cremmer-\-Scherk-\-Kalb-\-Ramond (CSKR) model coupled to scalars and fermions. The 2-form gauge potential
is proposed to couple non-minimally to matter, here described by a chiral scalar superfield.
The important outcome is that supersymmetry is kept exact
in the core and it may also hold in the exterior region of the string.
We contemplate the configurations of the bosonic sector and we check that the solutions
saturate the  Bogomol\`nyi bound. A glimpse on the fermionic zero
modes is also given.
\end{abstract}

\newpage

\section{Introduction}

High-Energy Physics and theories for fundamental interactions strongly
rely on the concept of
spontaneous symmetry breaking. In  Cosmology, the common belief
is that, at high temperatures, symmetries that are
spontaneously broken today have already been exact in the early stages of the Universe.
During the evolution of the Universe, there were various phase transitions, associated with
 the chain of spontaneous breakdowns of gauge symmetries.

Topological defects such as cosmic strings \cite{Vilenkin,Kibble1,Vilenkin1,Kibble2}, probably
produced during ''non-perfect symmetry-breaking transitions'' \cite{Kibble}
are  the object of our concern in this paper. Cosmic strings appear in some
Grand Unified Gauge
Theories and carry a large energy density \cite{Kibble1}. In this way, they may provide
a possible origin for the seed density perturbations which became the large
scale structure of the Universe observed today \cite{Stebbins,Sato}. These fluctuations would
leave their imprint in the cosmic microwave background radiation (CMBR) that
would act as seeds for structure formation and then as builders of the
large-scale structures in the Universe\cite{Brandenberger}. They may also
help to explain the most energetic events in the Universe such as high
energy cosmic rays\cite{Bhattacharjee1,Bhattacharjee2,Brandenberger2}.

In view of the possibility that Supersymmetry (su.sy) was realized in
the early Universe and was broken approximately at the same time as cosmic
strings were formed, many recent works investigate
cosmic strings by adopting a supersymmetric framework \cite{Morris,Davis}.

In the present work, we propose a possible model for  supersymmetric
cosmic strings. More specifically, we shall be dealing with consider the N=1-version of
the Cremmer-Scherk-Kalb-Ramond theory (CSKR). The anti-symmetric Kalb-Ramond
tensor field (KR) was first introduced
within a string theory context\cite{KR}. More generally, p-form gauge
fields appear in many  supergravity models \cite{Gates}.
In the cosmic string context, the KR-field is an important
ingredient to describe a global
$U(1)$-string \cite{Vilenkin2,Shellard,Davis3,Kaloper}, mainly because it brings about a massless scalar that is responsible for
a long-range attractive interaction.

To build up the supersymmetric extension of the CSKR model of our paper,  three superfields
are required; it has a rather simple field content and provides interesting results:
the Kalb-Ramond superfield is responsible for the  stablility of the potential, the vacuum
is supersymmetric, supersymmetry is not  broken in the string core, and
 gauge symmetry is broken without a Fayet-Iliopoulos term.
The interesting feature of our model is that su.sy  may be
kept exact, i. e. we may have a model where a genuinely
supersymmetric cosmic string appears, su.sy. being exact
in the string core and there is enough freedom to arrange
parameters in such a way that sy.sy. may also hold outside the string.

The outline of this paper is as follows: in Section 2, we start by
presenting the (CSKR) model in terms of superfields,  a component-field
description is also carried out and the on-shell version is finally
presented. In Section 3, we devote some time to calculate and discuss
the gauge-field propagators. Their poles are relevant in order to identify the
quanta that intermediate the interactions inside and
outside of the string.
Problems like gauge and supersymmetry breakings, field equations
and vortex configurations, as well the behaviour of the solutions inside and outside the
 defect, are the matter of Section 4. Section 5 is devoted to
 the study of the Bogomol$'$nyi duality
conditions. Finally, in Section 6, we draw our General Conclusions.

\section{Setting up a supersymmetric cosmic strings \zero}

In this section, we study the $N=1$-supersymmetric version of the
CSKR model
with non-minimal coupling to matter, envisaging the study of the formation of a cosmic string.

The ingredient superfields of the model are a chiral scalar supermultiplet, $\Phi $, the gauge
superpotential, ${\cal V}$, and a chiral spinor-valued superfield, ${\cal G}$ \cite{Nogue}.

The superspace action is given as follows

\begin{equation}
S = \int d^4x d^2\theta \left\{ -\frac{1}{8}{\cal W}^a{\cal W}_a + d^2 \bar
\theta \left[-\frac{1}{2}{\cal G}^2 + \frac{1}{2}m {\cal V G} + \frac{1}{16}
\bar \Phi e^{2h {\cal V}}\Phi e^{4g{\cal G}}\right]\right\} \label{action1}.
\end{equation}

\vspace{.4 true cm}

Below, we give the $\theta$-component expressions of the superfield
above, where  ${\cal V}$ is already assumed to be in the Wess-Zumino gauge:

\begin{equation}
\Phi = e^{-i\theta \sigma^{\mu} \bar \theta \partial_{\mu}}[\phi(x) +
\theta^{a}\chi_a(x) + \theta^2S(x)]; \label{sup1}
\end{equation}

\begin{equation}
{\cal V} = \theta \sigma^{\mu} \bar \theta A_{\mu}(x) + \theta^2 \bar \theta
\bar \lambda (x) + \bar\theta^2 \theta \lambda (x)+ \theta \bar \theta^2
\Delta (x); \label{sup2}
\end{equation}

\begin{equation}
\begin{array}{ll}
{\cal G} = & -\frac{1}{2} M + \frac{i}{4} \theta \xi_a - \frac{i}{4} \bar
\theta_{\dot a} \bar \xi^{\dot a} + \frac{1}{2} \theta \sigma^{\mu}_{a\dot
a} \bar \theta ^{\dot a} \tilde G_{\mu} + \frac{1}{8} \theta^a
\sigma^{\mu}_{a \dot a} \bar \theta^2 \partial_{\mu} \bar \xi^{\dot a} + \\
& - \frac{1}{8} \theta^2 \sigma^{\mu}_{a \dot a} \bar \theta^{\dot
a}\partial_{\mu} \xi^a - \frac{1}{8} \theta^2 \bar \theta^2 \Box M;
\end{array}\label{sup3}
\end{equation}

\noindent
with the superfield-strenght $ {\cal W}_a $  written as

\begin{equation}
{\cal W}_a = -\frac{1}{4} \bar D^2 D^a {\cal V}.
\end{equation}

\noindent
$\tilde G_{\mu}$ is the dual of the $3$-form field strength,
$G_{\mu \nu k}$, related to the $2$-form Kalb-Ramond (KR) field, $B_{\mu \nu}$:

\vspace{.3 true cm}

\begin{equation}
G_{\alpha \mu \nu} = \partial_{\alpha} B_{\mu \nu} + \partial_{\mu} B_{\nu
\alpha} + \partial_{\nu} B_{\alpha \mu},
\end{equation}

\vspace{.3 true cm}

\begin{equation}
\tilde G_{\mu} = \frac{1}{3 !}\epsilon_{\mu \nu \alpha \beta} G^{\nu \alpha \beta}.
\end{equation}

\vspace{.3 true cm}

\noindent
It is worthehile to mention that, though the superfield ${\cal V}$ (where $A^{\mu}$
is accommodated) appears  explicitly in the action of the model, the $B_{\mu \nu}$-field
 shows up only through is field-strength, located in the chiral scalar superfield ${\cal G}$.
 As for the $\chi$, $\lambda$ and $\xi$, these are the fermionic partners of the scalar
 matter, the photon and the KR gauge potential, respectively.
Our conventions for the supersymmetry (su.sy.) covariant derivatives are  given  as below:

\begin{equation}
\begin{array}{ll}
D_a = \partial_a - i \sigma^{\mu}_{a \dot a} \bar \theta^{\dot a}
\partial_{\mu}, &  \\
\bar D_{\dot a} =- \partial_{\dot a} + i \theta^a \sigma^{\mu}_{\dot a a}
\partial_{\mu}, &
\end{array}
\end{equation}

\noindent
where the $\sigma^{\mu}$-matrices read as $\sigma^{\mu} \equiv ({\bf 1}; \sigma^i)$, the $\sigma^i$'s
being the Pauli matrices.

For a detailed description of the component fields and degrees of freedom
 displayed in (\ref{sup1}-\ref{sup3}), the
reader is referred to the paper of ref. \cite{Nogue}.

As it can be readily checked, this action is invariant under
two independent sets of Abelian gauge transformations, with parameters

\begin{equation}
\begin{array}{llll}
\phi(x) &\rightarrow & \phi'(x)= \phi(x)e^{i\Lambda (x)}, &\\
A_\mu(x) &\rightarrow & A^{\prime}_\mu(x) = A_\mu (x) - \partial_\mu \Lambda(x),
&  \\
B_{\mu \nu}(x) &\rightarrow &B'_{\mu \nu}(x) = B_{\mu \nu}(x) +
\partial_{\mu}\xi_{\nu}(x) - \partial_{\nu}\xi_{\mu}(x), &
\end{array}
\end{equation}

\vspace{.3 true cm}

\noindent
where the arbitrary functions $\Lambda $ and $\xi_{\mu}$ vanish at
infinity. The topological term present in (\ref{action1}),
$m {\cal V}{\cal G}  $, will play a crucial role in the analysis
of the breaking of supersymmetry,
as we shall find out in next section.

When we write the action in terms of components fields, we
obtain the following equations for  the auxiliary fields:

\begin{eqnarray}
S^* + \frac{i g}{2} \bar X \Gamma_R \Xi + \frac{g^2}{4}\bar \Xi \Gamma_R \Xi
\phi^* &=&0; \label{S}\\
\Delta + \frac{1}{2}\left[he^{-2gM}\phi \phi^*\right] - 2mM&=&0.\label{D}
\end{eqnarray}

\vspace{.3 true cm}

\noindent
where $X$, $\Xi$ and $\Lambda$ are 4-component spinors so defined that:

\begin{equation}
\Xi \equiv \left(
\begin{array}{ll}
\xi_a(x)  \\
\bar \xi^{\dot a}(x)
\end{array}
\right), \hspace{.3 true cm} X \equiv \left(
\begin{array}{ll}
\chi_a(x)   \\
\bar \chi^{\dot a}(x)
\end{array}
\right), \hspace{.3 true cm} \Lambda \equiv \left(
\begin{array}{ll}
\lambda_a(x)   \\
\bar \lambda^{\dot a}(x)
\end{array}
\right); \label{ferm}
\end{equation}

\vspace{.3 true cm}

\noindent
in this $4$-dimensional representation, the Dirac matrices are given as below:

\begin{equation}
\Gamma^{\mu}=\left(\begin{array}{cc}
0& \sigma^{\mu}_{ab}\\
\bar \sigma^{\mu \dot a b}&0
\end{array}\right), \hspace{.4 true cm} \Gamma_5 \equiv i \Gamma^0 \Gamma^1 \Gamma^2 \Gamma^3.
\end{equation}

Eq.(\ref{ferm}) is simply a rewriting of the Majorana spinors, $\Xi$, $\chi$ and $\Lambda$, in
terms of their chiral (L- and R- handed ) components.

Using the equation of motion for $\Delta$, we have:

\begin{equation}
U = \frac{1}{2}\left[h |\phi|^2 e^{-2gM} - 2mM\right]^2 \geq 0 \label{pot}.
\end{equation}

\noindent
This potential shall be thoroughly discussed in the next section,
in connection with the study of the breakdown of the $U(1)$-symmetry
supersymmetry in the core and outside the defect.
Notice that $U$ is non-negative, as expected by virtue of supersymmetry.

Using (\ref{S}) and (\ref{D}), we next  eliminate $S$ and $\Delta $ from the
component-field action and rewrite  the Lagrangian density exclusively in terms
of the physical component fields. The Lagrangian
takes now the form:

\begin{equation}
{\cal L} = {\cal L}_B + {\cal L}_F + {\cal L}_Y - U,
\end{equation}

\noindent
where the bosonic part, $ {\cal L}_B$, reads:

\begin{equation}
\begin{array}{ll}
{\cal L}_B =& \nabla_{\mu} \phi [\nabla^{\mu} \phi]^* e^{-2gM} + {\cal P}\partial_{\mu}M \partial^{\mu}M-\frac{1}{4}F_{\mu \nu}F^{\mu \nu} \\
&  + \frac{1}{6} G_{\mu \nu k}
G^{\mu \nu k}  +  m \epsilon^{\mu \nu \alpha \beta} A_{\mu}
\partial_{\nu}B_{\alpha \beta},
\end{array}\label{lag1}
\end{equation}

\noindent
with

\begin{equation}
{\cal P} = 1-g^2|\phi|^2 e^{-2gM} \label{P}.
\end{equation}

\noindent
The kinetic Lagrangian for the fermionic sector is given by

\begin{equation}
{\cal L}_F = \frac{i}{2} \bar
\Lambda \Gamma^{\mu} \partial_{\mu} \Lambda + \frac{i}{4}{\cal P} \Xi \Gamma^{\mu} \partial_{\mu} \Xi +
\frac{i}{4}\bar X \Gamma^{\mu} \nabla_{\mu 5} X e^{-2gM},
\end{equation}

\noindent
whereas the Yukawa Lagrangian is given as below:

\begin{equation}
\begin{array}{ll}
{\cal L}_Y = &i {\cal N} \bar \Lambda \Gamma_5 \Xi + \left[- h(\phi \bar \Lambda \Gamma_R X +
\phi^* \bar \Lambda \Gamma_L X)- \frac{g^2}{4h} \bar \Xi \Gamma_5 \Gamma^{\mu} {\cal J}_{\mu}e^{2gM} \Xi \right.\\
 &+ \left. - \frac{g^2}{2} \partial_{\mu} M
(\bar X \Gamma_{L} \Gamma^{\mu} \Xi \phi^* + \bar \Xi \Gamma_{L} \Gamma^{\mu} X \phi )+ \frac{g}{2} ( \Xi \Gamma^{\mu} \Gamma_R X \nabla_{\mu} \phi  \right.\\
 &\left.+\bar X \Gamma_L \Gamma^{\mu} \Xi [\nabla_{\mu} \phi]^*)  \right]e^{-2gM},
\end{array}
\end{equation}

\noindent
where

\begin{equation}
{\cal N} = m + gh |\phi|^2e^{-2gM}.
\end{equation}

\noindent
and $\Gamma_{L,R}$ are the left- and right-sector projectors, respectively.

\noindent
The current ${\cal J}_{\mu}$ that appears in ${\cal L}_Y$ can be expressed according to:

\begin{equation}
{\cal J}_{\mu} = -\frac{ih}{2} \left( \phi^* \nabla_{\mu } \phi -
\phi [\nabla_{\mu} \phi]^*\right)e^{-2gM},  \label{curr1}
\end{equation}

\noindent
with

\begin{equation}
\nabla_{\mu} \phi = \left(\partial_\mu + ihA_\mu + ig\tilde G_{\mu}
\right)\phi.
\end{equation}

\noindent
At last, the covariant derivative with $\Gamma _5$-couplings is given by

\begin{equation}
\nabla _{\mu 5}X=\left( \partial _\mu -ihA_\mu \Gamma _5-ig\tilde{G}_\mu
\Gamma _5\right) X.
\end{equation}

The topological current, namely, the one whose divergencelessness
follows without any reference to equations of motion or to symmetries is given by

\begin{equation}
K^{\mu} = \partial_{\nu}(\tilde F^{\mu \nu} + \tilde B^{\mu \nu}).
\end{equation}

\noindent
However, due to the absence of magnetic monopoles the first term can be thrown away and we get

\begin{equation}
K^{\mu} = \partial_{\nu} \tilde B^{\mu \nu},
\end{equation}

\begin{equation}
\partial_{\mu} K^{\mu} \equiv 0.
\end{equation}

\noindent
The corresponding topological charge is

\begin{equation}
Q\equiv \int K^0 d^3x = \int d^3x {\cal B},
\end{equation}

\noindent
where ${\cal B}$ stands for the magnetic-like field (a scalar)
associated to the 2-form potential, which is known to be the divergence of a
vector potential:

\begin{equation}
{\cal B} = \vec \nabla . \vec b.
\end{equation}

The model is now completely set up. Next, we shall
discuss the spectrum of excitations in order
to get relevant information on the interactions mediated
by the gauge particles.

\section{Propagators and Interaction Range. \zero}

The aim of this section is to compute the propagators for the
gauge-field excitations. For the sake of generality, we shall
suppose that both scalars, $ \phi $ and $ M $, acquire non-vanishing
vacuum expectation values, $<\phi> = \eta$
and $<M> = \rho $. Whether or not $\eta$ and $\rho$ vanishing will be
seen when we will study the minimum of the potential and
we will discuss the spontaneous breaking of the symmetries (Section 4).

For this purpose, we parametrize $ \phi $  as below:

\begin{equation}
\phi =[\phi (x)^{\prime }+\eta ]e^{i\Sigma (x)},
\end{equation}

\noindent
where $ \phi ^{\prime } $ is the quantum fluctuation around the ground state $\eta$.

To analyse the propagators, we need to concentrate on the bosonic Lagrangian in terms
of the physical fields $\phi ^{\prime } $,  $A_\mu $ and $B_{\mu \nu }$.
For the sake of reading off these propagators, we refer to the
bilinear sector of the bosonic Lagrangian, whose kinetic piece can be cast like below:

\begin{equation}
\begin{array}{ll}
{\cal L}_K = &\frac{1}{4} F_{\mu \nu} F^{\mu \nu} -
(1-g^2 \zeta \eta^2) \tilde G_{\mu} \tilde G^{\mu} +
2(m + g h \zeta \eta^2) A_{\mu}\tilde G^{\mu} +\\
&- 2h\zeta \eta^2 \Sigma \partial_{\mu} A^{\mu} + h^2 \zeta \eta^2 A_{\mu} A^{\mu} +
\zeta \eta^2 \partial_{\mu} \Sigma \partial^{\mu} \Sigma , \label{kin}
\end{array}
\end{equation}

\noindent
where $\zeta \equiv e^{-2g \rho } = e^{-2g<M>} $.

At first, we will write it in a more convenient form:

\begin{equation}
{\cal L} = \frac{1}{2}\sum_{\alpha \beta} {\cal A}^{\alpha} {\cal O}_{\alpha
\beta} {\cal A}^{\beta},
\end{equation}

\vspace{0.3cm}

\noindent
where ${\cal A} _\alpha =(\Sigma ,A_\mu ,B_{\mu \nu })$ and ${\cal O}_{\alpha
\beta }$ is the wave operator. We notice that $\Sigma$ mixes with $A^{\mu}$. However, if
we adopt the t'Hooft $R_{\xi}$-gauge, they decouple from each other. So,
the $\Sigma-\Sigma$ propagator can be derived independently from the propagators for the
($A^{\mu} ,B^{\mu \nu}$) sector.

In order to invert the wave operator, we have to fix the gauge so as to make the matrix
 non-singular. This is accomplished by adding the gauge-fixing terms,

\begin{eqnarray}
{\cal L}_{A_\mu } &=&\frac 1{2\alpha }(\partial _\mu A^\mu + \alpha \zeta h^2 \eta^2 \Sigma
)^2; \\[0.3cm]
{\cal L}_{B_{\mu \nu }} &=&\frac 1{2\beta }(\partial _\mu B^{\mu \nu
})^2.
\end{eqnarray}

\vspace{0.3cm}

To read off the gauge-field propagators, we shall use an extension of the spin-projection operator formalism
presented in \cite{Rivers,Nitsch}. In this work, we have to add other new
operators coming from the Kalb-Ramond terms. The two operators which act on
the tensor field are:

\begin{eqnarray}
(P^1_b)_{\mu \nu, \rho \sigma} &=& \frac{1}{2} (\Theta_{\mu \rho}\Theta_{\nu
\sigma} - \Theta_{\mu \sigma}\Theta_{\nu \rho}), \\
[0.3cm] (P^1_e)_{\mu \nu, \rho \sigma} &=& \frac{1}{2} (\Theta_{\mu
\rho}\Omega_{\nu \sigma} - \Theta_{\mu \sigma}\Omega_{\nu \rho} -\Theta_{\nu
\rho}\Omega_{\mu \sigma} + \Theta_{\nu \sigma}\Omega_{\mu \rho}),
\end{eqnarray}

\noindent
where $\Theta_{\mu \nu}$ and $\Omega_{\mu \nu}$ are, respectively, the
transverse and longitudinal projection operators, given by:

\begin{equation}
\Theta _{\mu \nu }=\eta _{\mu \nu }-\Omega _{\mu \nu },
\end{equation}

\noindent
and

\begin{equation}
\Omega_{\mu \nu} = \frac{\partial_{\mu} \partial_{\nu}}{\Box }.
\end{equation}

\vspace{0.3cm}

\noindent
The other operator coming from the Kalb-Ramond sector, $S_{\mu \gamma k}$, is
defined in terms of Levi-Civita tensor as

\begin{equation}
S_{\mu \gamma k} = \epsilon_{\lambda \mu \gamma k} \partial^{\lambda}.
\end{equation}

\vspace{.3 true cm}

In order to find the wave operator$'$s inverse, let us calculate the
products of operators for all non-trivial combinations involving the
projectors. The relevant multiplication rules are as follows:

\vspace{0.4cm}
\begin{equation}
\begin{array}{ll}
(P^1_b)_{\alpha \beta,\lambda \xi} S^{\lambda \xi \nu} = 2\epsilon_{\rho
\alpha k}^{\hspace{.5 true cm}\nu}\partial^{\rho}; &  \\
&  \\
S_{\alpha \lambda \rho}(P^1_b)^{\lambda \rho, \xi \eta}= 2\epsilon_{\rho
\alpha}^{\hspace{.4 true cm}\xi \eta} \partial^{\rho}; & \\
&  \\
(P^1_e)_{\alpha \beta,\lambda \xi} S^{\lambda \xi \nu} =0; &  \\
&  \\
S_{\alpha \lambda \rho}(P^1_e)^{\lambda \rho, \xi \eta}=0. &
\end{array}
\end{equation}

\vspace{0.3cm}

\noindent
The wave operator ${\cal O}$ can be split into four sectors, according to:

\vspace{.3 true cm}

\begin{equation}
{\cal O} = \left(
\begin{array}{ll}
A & B \\
C & D
\end{array}
\right),
\end{equation}

\vspace{.3 true cm}

\noindent with

\begin{eqnarray}
A &=&M_1\Theta _{\mu \nu }+M_2\Omega _{\mu \nu }; \\[0.3cm]
B &=&-M_3S_{\mu \nu k}; \\[0.3cm]
C &=&M_3S_{\alpha \beta \nu }; \\[0.3cm]
D &=&M_4(P_b^1)_{\alpha \beta ,\nu k}-M_5(P_e^1)_{\alpha \beta ,\nu k},
\end{eqnarray}

\vspace{0.4cm}

where $M_1,...,M_5$ are  quantities that read as the following expressions:

\begin{equation}
\begin{array}{ll}
M_1= & \frac{1}{2}(\Box + 2h^2 \zeta \eta^2), \\
\\
M_2= & - \frac{1}{2}(\frac{1}{\alpha}\Box - 2h^2 \zeta \eta^2), \\
\\
M_3= & \frac{1}{2}(m + hg \zeta \eta^2), \\
\\
M_4= & \frac{1}{2}(1-g^2\zeta \eta^2) \Box , \\
\\
M_5= & \frac{1}{8\beta}\Box ,
\end{array}
\end{equation}

\noindent
$\zeta $ was defined in the Lagrangian (\ref{kin}).

After lengthy algebraic calculations, we are able to read off the operator ${\cal O}^{-1}$.

\vspace{.3 true cm}

\begin{equation}
{\cal O}^{-1}=\left(
\begin{array}{ll}
X & Y \\
Z & W
\end{array}
\right) ,
\end{equation}

\vspace{.3 true cm}

\noindent where the quantities $X$, $Y$, $Z$ and $W$ are:

\begin{equation}
\begin{array}{ll}
X = (A - B D^{-1} C)^{-1}, &  \\
Z=-D^{-1}CX, &  \\
W=(D-CA^{-1}B)^{-1}, &  \\
Y = - A^{-1}BW, &
\end{array}
\end{equation}

\vspace{0.3cm}

\noindent
or, explicitly, in terms of matrix elements,

\begin{equation}
<AA> = \frac{i}{(M_1 + 4 M_3^2M_4^{-1}\Box)} \Theta_{\mu \nu} +
\frac{i}{M_2}
\Omega_{\mu \nu};  \label{propag1}
\end{equation}

\begin{equation}
<AB> = -2 M_3 M_4^{-1}[M_1 + 4 M_3^2
M_4^{-1}\Box]^{-1}
\epsilon^{\lambda \mu \nu }_{\alpha} \partial_{\lambda} \Theta ^{\alpha k}
\end{equation}

\begin{equation}
<BB> = \frac{i}{M_4M^{-1}_1(M_1 + 4M_3^2M_4^{-1}\Box )}(P^1_b)_{\alpha \beta \gamma k } +
\frac{i}{(4M_3^{-1}M_1^{-1}\Box - M_5 )}(P_e^1)_{\alpha \beta \gamma k}.
\label{propag2}
\end{equation}

\noindent
As for the $\Sigma - \Sigma $ propagator, it comes out as

\begin{equation}
<\Sigma \Sigma> = -[\alpha \zeta \eta^2(\frac{1}{\alpha}\Box -
2 h^2 \zeta \eta^2)]^{-1} = [2 \alpha \zeta \eta^2 M_2]^{-1}
\end{equation}

\noindent
The poles of the transversal part of the $A_{\mu}$- and $B_{\mu \nu}$-
propagators (\ref{propag1})-(\ref{propag2}) are  the same and given by

\begin{equation}
k^2= 2h^2 \zeta \eta^2 + 8 \frac{(m + hg\zeta \eta^2)^2}{(1-g^2 \zeta \eta^2)}.
\end{equation}

Therefore, $A_{\mu }$ and $B_{\mu \nu }$ no longer mediate long- range interactions:
their physical degrees of freedom combine into physical massive quanta responsible
for the short-range character of the interaction. Next, we shall use these results
to infer about a vortex-like configuration leading to a particular type of cosmic string.

\section{The  Cosmic-String Field Configuration \zero}

We will finally analyse the possibility of obtaining the
cosmic string   based on the bosonic Lagrangian  eq.(\ref{lag1}).

The vacuum of the theory is obtained by analysing the minimum of the potential
given by eq.(\ref{pot}).
This system has an extremum at $|\phi | = 0$. This extremum corresponds
to an unstable one and, of course, does not provide cosmic string formation.
The  minimum is set by the equation $\rho = \frac{h \eta ^2 \zeta }{2 m} $.

The minimization of our potential reveals an interesting feature: in the
string core, neither gauge symmetry nor supersymmetry are
broken; on the other hand, outside  the
defect,  gauge symmetry is spontaneously broken by $<\phi>$, while su.sy. is still kept exact.

In the string core: $<\phi > = <M>=0 $, so no breaking takes place. Outside the string core:
$<\phi > = \eta \neq 0$, $<M> = \rho \neq 0$, $U = 0$; only gauge symmetry is broken.

This might be of interest if we keep in mind the possibility that cosmic strings might
have been formed before the phase transition leading to su.sy. breaking.
This means that su.sy at the string core is an inheritance of the su.sy. era.

This model has a cosmic string solution with
vortex configuration,

\begin{eqnarray}
\phi  &=&\varphi (r)e^{i\theta },  \nonumber \\[0.3cm]
A_\mu  &=&\frac 1h(A(r)-1)\delta _\mu ^\theta .  \label{vortex1}
\end{eqnarray}

\vspace{0.3cm}

\noindent
This configuration has the same form of ordinary cosmic string \cite{Nielsen},
parametrized in cylindrical coordinates $(t,r,\theta , z)$, where $r \geq 0$
and $0\leq \theta < 2 \pi $. The boundary conditions for the fields $\varphi $
and $A( r )$ are

\begin{equation}
\begin{array}{ll}
\varphi ( r ) = \eta & r \rightarrow \infty \\
\varphi ( r ) = 0 & r =0
\end{array}
\hspace{.6 true cm}
\begin{array}{ll}
A( r ) = 0 & r \rightarrow \infty; \\
A( r ) = 1 & r = 0.
\end{array}
\label{cvortex1}
\end{equation}

\vspace{0.3cm}

\noindent
We propose the following vacuum configuration for the $M$-field:

\begin{equation}
M = M(r),  \label{vortex2}
\end{equation}

\noindent
with the boundary conditions below:

\begin{equation}
\begin{array}{ll}
M ( r ) = \rho & r \rightarrow \infty; \\
M ( r ) = 0 & r = 0.
\end{array}
\label{cvortex2}
\end{equation}

The vortex configuration for the dual field, $\tilde{G}_\mu $, which preserves
the cylindrical symmetry is written as

\begin{equation}
\tilde G_{\mu } = \frac{G(r)}{g r} \delta^{\theta}_{\mu} ,  \label{vortex3}
\end{equation}

\noindent
with boundary conditions given by:

\begin{equation}
\begin{array}{ll}
G(r)=0 & r\rightarrow \infty;  \\
G(r)=0  & r=0.
\end{array}
\label{cvortex3}
\end{equation}

\noindent
The analysis carried out for the propagators leads us to propose that
the $G$-field field does not have propagation outside  the string.
It is sensible to propose that its only
non-vanishing component is the angular one ($\theta -$component ) , because
this is the only one that couples to the gauge field, $A_{\mu}$.

The Euler-Lagrange equations stemming from eq.( \ref{lag1}), combined with
the vortex  conditions by Nielsen-Olesen-\cite{Nielsen}, when applied to the
$\varphi$ field give us

\begin{equation}
\frac{1}{r}\frac{\partial}{\partial r} \left(r\frac{\partial \varphi}{\partial r}\right)  -
h^2 \varphi^3e^{-2gM} + 2h m M \varphi - \frac{{\cal H}^2 \varphi}{r^2} - 2g M'\varphi' + g^2 M'^{2} \varphi =0,
\label{mov1}
\end{equation}

\noindent
where the field strenght, ${\cal H}$, is defined by

\begin{equation}
{\cal H}^2 =   A^2 + G^2 + 2AG.
\end{equation}

The dynamics of the gauge field, $A_{\mu}$, with the vortex configuration,
eq.(\ref{vortex1}), is
given by

\vspace{0.3cm}

\begin{equation}
r\frac{\partial}{\partial r} \left(\frac{1}{r}\frac{\partial A}{\partial r}\right)
 - \frac{2m}{g}h G - 2 h^2[A \varphi^2
 + G  \varphi^2]e^{-2gM} = 0.\label{mov3}
\end{equation}

\noindent
These equations
have a different form from ordinary cosmic string, but are compatible
with its vortex configuration.

The dynamics of the $M$-field   is governed by:

\vspace{0.3cm}

\begin{equation}
\frac{1}{r}{\cal P}(rM')' - 2g^2 \varphi' M'e^{2gM} + g{\cal X}^2  -g\frac{{\cal H}^2}{r^2}\varphi^2e^{-2gM} - {\cal Y} = 0,\label{mov2}
\end{equation}

\vspace{0.3cm}

\noindent
where we define

\begin{equation}
\begin{array}{ll}
{\cal X}^2 = & g^2 \varphi ^2M'^2     + \varphi'^2e^{2gM} ;\\
{\cal Y}=& -( hm\varphi^2 + 2ghmM\varphi^2 ) e^{-2gM}+ 2 m^2 M -  g h^2\varphi^4 e^{-4gM}.
\end{array}
\end{equation}

\noindent
The term labeled by ${\cal Y}$ comes from the derivative of the potential.

The equation of motion for the tensorial field, $B_{\mu \nu}$, can
be written in terms of the dual, $\tilde G_{\mu}$, as

\begin{equation}
[G - \frac{mg}{h}A -  g^2\varphi^2\left(A  +  G\right)e^{-2gM}]'  = 0 ;\label{mov5}
\end{equation}

\vspace{0.3cm}

We point out that the $M$-and-$G$-fields are essential in this
formulation of ordinary cosmic string. As shown, the $M$-field
appearing in the eq.(\ref{pot}) for the potential is responsible for
its shape and plays a key r\^ole for the non-breaking of supersymmetry.

This matter deserves a better discussion. The fact that the potential
$U$ is non-negative is a consequence of supersymmetry. However, to
be sure that supersymmetry is actually not broken, one must check that $U$
is indeed vanishing for the configuration under consideration. Actually, we can see
that, in the string core, by virtue of the boundary conditions
(\ref{cvortex1}),(\ref{cvortex2}) and (\ref{cvortex3}), $U$ is
zero and so supersymmetry holds true in the interior region.
Outside the core, $\rho $ must be so chosen that $U=0$,
if supersymmetry is to be kept exact; namely, $\rho$ and $\eta $
are related to one another by $2m\rho = h \eta^2 e^{-2g \rho}$.
With this relationship, which is a sort of fine-tunning, we can
keep supersymmetry also outside the string. So, we conclude that our
model is able to accommodate a cosmic string configuration compatible
with $N=1$-supersymmetry. On the other hand, if we are to reproduce
the more realistic situation with broken su.sy. outside  the string,
we simply  do not take $\rho$ and $\eta$ as related by the
fine-tunning we alluded to above. In such a case, we have su.sy.
inside the core, as a relic of the string formation in the su.sy era,
and no su.sy. outside, in agreement with current models and
phenomenology \cite{Davis,Morris}.

At this stage, some highlights on the fermionic zero modes are in order.
This issue shall be discussed with a great deal of details in a forthcoming
paper \cite{Cris2}; however, we may already quote some
preliminary results. All we have done previously concerns the bosonic sector
of the $N=1$- CSKR theory; to introduce the fermionic modes which have partnership
 with the configurations (\ref{vortex1}),(\ref{vortex2}) and (\ref{vortex3}),
take advantage from su.sy invariance. By this, we mean that the configurations
for the fermionic degrees of freedom may be found
out by acting with su.sy. transformations on the bosonic sector. In a paper by Davis
et al. \cite{Davis}, this procedure is clearly stated  and we follow it here.
The su.sy. transformations of the component fields for the
$N=1$-CSKR model may
be found in the work of ref.\cite{Nogue}

Going along the steps of ref. \cite{Nogue}, we quote below the fermionic
configuration, we have worked out:

\begin{equation}
\chi_a = 2i \bar \varepsilon ^{\dot a}e^{i\theta}\left[\sigma^1_{a \dot a}\varphi' +
i \sigma^2_{a \dot a}A\varphi \right],\label{ferm1}
\end{equation}

\begin{equation}
\xi _a = 2 \bar \varepsilon^{\dot a} \left[\sigma^1_{a \dot a}M' -
\frac{i}{gr}\sigma^2_{a \dot a} G \right],\label{ferm2}
\end{equation}

\begin{equation}
\lambda _a = \varepsilon _{a}\left[2mM - \frac{1}{2}h\varphi^2e^{-2gM}\right] -
i \varepsilon^b \sigma^1_{b \dot a}\sigma^{2 \dot a}_a A',
\label{ferm3}
\end{equation}

\noindent
where $\sigma^{1,2}$ refer to $r, theta$-components in cylindrical
coordinates.

It is worthwhile to notice that the su.sy. transformations lead to  a vector supermultiplet ($\cal V $) that
is no longer in the Wess-Zumino gauge; to reset such a gauge for $\cal V$, we
have to supplement the su.sy transformation by a suitable gauge transformation
that has to act upon the matter fields as well. $\cal G$ is gauge invariante, so this further
gauge transformation should be carried out only on $\Phi$ and $\cal V $; the results presented
in eqs.(\ref{ferm1})- (\ref{ferm3})
take already into account the c

ombined action of a su.sy. and a gauge transformation.

The equations of motion above do not have an exact solution;
they can may be better handled by considering the Bogomol$'$nyi
limit \cite{Linet}. We show that in
the supersymmetric CSKR theory, we can
find the limit where the equations of motion can be
written as first order equations. In the next section,
we will show that the $M$-and-$G$-fields may be related
to each other by means of the  Bogomol$'$nyi equations.

\section{Bogomol$'$nyi limit \zero}

In this section, we show that it is possible to find the
energy-momentum tensor for a thin cosmic string in the CSKR theory,
in the flat space and we analyse  the
possibility of to find a Bogomol$'$nyi solution compatible with the Bogomol$'$nyi bound.
The energy-momentum tensor is defined as

\begin{equation}
T^{\mu}_{\nu} = 2 g^{\mu \alpha}\frac{\partial {\cal L}}{\partial
g^{\alpha \nu}} - \delta^{\mu}_{\nu}  {\cal L}.
\end{equation}

\noindent
The non-vanishing components of the energy-momentum tensor are

\begin{equation}
T^t_t = T^z_z = \varphi'^2 e^{-2gM}+\frac{A^2\varphi^2 }{r^2}e^{-2gM} + {\cal P} M'^2 +
\left(\frac{A'}{r h}\right)^2 + {\cal R}\left(\frac{ G}{gr} \right)^2 + U.
\label{Tt}
\end{equation}

\noindent
From (\ref{Tt}) we see that the boost symmetry is not break. This
is related to the absence of current  in
$z$-direction, as in the usual ordinary
cosmic string.

The transverse components of the energy-momentum tensor are given by:

\begin{equation}
T^r_r = -\varphi'^2 e^{-2gM} +\frac{A^2\varphi^2 }{r^2}e^{-2gM}-
{\cal P} M'^2 - \left(\frac{A'}{r h}\right)^2 +
{\cal R}\left(\frac{ G}{gr} \right)^2 + U ;
\end{equation}

\begin{equation}
T^{\theta}_{\theta} = \varphi'^2 e^{-2gM} -
\frac{A^2\varphi^2 }{r^2}e^{-2gM}+ {\cal P} M'^2 -
\left(\frac{A'}{r h}\right)^2 - {\cal R}\left(\frac{ G}{gr} \right)^2  + U ,
\end{equation}

\noindent
where ${\cal P}$ is the same that was defined in
eq.(\ref{P}), and ${\cal R}$ is

\begin{equation}
{ \cal R}(r) = 1+ g^2\varphi^2 e^{-2gM}.
\end{equation}

\noindent
This  energy-momentum tensor has only an $r$-dependence.

We define the energy density per unit of length, ${\cal E}$

\begin{equation}
{\cal E} = 2\pi \int_0^{\infty} T_t^t r dr,
\end{equation}

\noindent
with $T^t_t$ given by Eq.(\ref{Tt}). This can be written as

\begin{equation}
\begin{array}{ll}
{\cal E} =& 2 \pi  \int r dr\left[ \frac{1}{2}\left(\frac{A'}{rh} - h \varphi^2 e^{-2gM} + 2mM\right)^2 +
\frac{1}{r}\varphi^2 e^{-2mM} A' - \frac{2}{rh}m M A' + \right.\\
& \left.+ \left(\varphi' - \frac{A \varphi}{r}\right)^2e^{-2gM} +
\frac{2}{r}\varphi \varphi' A e^{-2gM} + \left(M' +
\frac{G}{gr}\right)^2 - \frac{2}{g r} G M' \right];
\end{array}
\end{equation}

\noindent
and the energy is given by

\begin{equation}
{\cal E} = 2 \pi \int r dr \left[ \frac{1}{r}\left( \varphi^2 A \right)'
e^{-2gM} - \frac{2m}{r h}A'M - \frac{2}{r}G M' \right].
\end{equation}

\noindent
Integrating by parts, and using the boundary conditions,  we find

\begin{equation}
{\cal E} \geq 8\pi m \rho/h.
\end{equation}

In this Bogomonl$'$nyi limit, there is not forces between vortices and
the equations of motion are of first order. The
search for  a Bogomol$'$nyi bound for the energy
yields the following system
of equations:

\begin{equation}
\begin{array}{ll}
\varphi'- \frac{A\varphi }{r} =0\\
\frac{A'}{rh} - h\varphi^2 e^{-2gM} + 2mM \\
M'+ \frac{G}{g r} =0,
\end{array}
\end{equation}

\noindent
with G given by eq.(\ref{mov3}).  With these equations, the transversal
components of the energy-momentum tensor density vanish and we get in the sense
of the distribution:

\begin{equation}
T^{\mu}_{\nu} = {\cal E} diag(1,0,0,1) \delta(x)\delta(y)
\end{equation}

This analysis provides a complete description of bosonic cosmic
strings and we realize
that is possible to find a consistent energy-momentum tensor on the
bosonic sector.

\section{Conclusions}

In this work, we have shown that we  may obtain a cosmic string in a supersymmetric
scenario  with the peculiarity that
supersymmetry breaking may not take place. The fact that susy might
(under certain conditions) be kept exact in the CSKR theory is mainly due to
the presence of  the $M$-field, partner of the $B_{\mu \nu}$-field,
introduced in the Kalb-Ramond superfield.

Introducing gravity in the model opens up a series of
interesting questions to tackle. Gravity by itself is
a relevant matter in connection
with cosmic string. In our model, once we are dealing with supersymmetry,
to bring about gravity we must consider the coupling of the  action (\ref{action1}) to
supergravity. $N=1-$matter coupling to local su.sy. \cite{Scherk,Ferrara,West,Julia}
introduces some arbitrary functions into the model; these functions
depend on the scalar matter.
It would be a pertinent question to try to understand whether (and how) these functions
(usually, referred to in the literature as $d$ and $g$) may be restricted
by the condition of string formation. This would allow us to restrict the universe of
d- and $g$ functions guided by the condition that cosmic strings form up in
the presence of gravity.

Another point that we would like to address to  concerns the topological charge of our solution.
>From the analysis of the propagators (Section 3), we clearly see that the correlation
functions for the $A^{\mu}$-and-$B^{\mu \nu}$-fields falls off outside the string.
Since the topological charge, Q, gets contribution exclusively from the magnetic scalar,
$\cal B$, (which is the divergence of the vector potential, $\vec b \sim B^{0i}$),
it readily follows that $Q$ vanishes, for it only feels the Kalb-Ramond potential,  $\vec b$,
at
infinity. The stability of the solutions is ensured by the topological charge conservation.
We have a solution that lives in the $Q=0$-sector and it is therefore stable.

Finally, an interesting   point we should analyse further concerns the property
of superconductivity of the string and the consideration of a
non-trivial fermionic background. A non-trivial fermionic background is relevant
in connection with supersymmetry
breaking. So, by  coupling to supergravity, a spontaneous su.sy. breaking. might be
induced and a careful analysis of the fermion background would be an
interesting issue to be contempleted.

\vspace{1 true cm}

{\bf Acknowledgments:}
The authors would like to express their deep gratitude to
Prof R. H. Brandenberger for helpful discussions  on the subject of this paper.
C.N F. would like to thank (CNPq-Brasil) and
M.B.S.M.P. would like to thank (FAPERJ-Rio de Janeiro) for financial
support. We are also grateful to the
thank Centro Brasileiro de Pesquisas F\'{\i}sicas (CBPF-CCP) and
Universidade Cat\'olica de Petr\'opolis (UCP-GFT) for the kind hospitality
during the preparation of this work.

\end{document}